# Image cloning beyond diffraction based on coherent population trapping in a hot Rubidium vapor

Dong-Sheng Ding[1,2], Zhi-Yuan Zhou[1,2], Bao-Sen Shi[1,2,*]

[1]*Key Laboratory of Quantum Information, University of Science and Technology of China, CAS, Hefei, Anhui 230026, China*
[2]*Synergetic Innovation Center of Quantum Information & Quantum Physics, University of Science and Technology of China, Hefei, Anhui 230026, China*
[*]*Corresponding author: drshi@ustc.edu.cn*

Following the recent theoretical predictions given in a paper [PRA 88, 013810 (2013)], we reported on an experimental realization of an image cloning beyond usual diffraction through coherent population trapping (CPT) effect in a hot rubidium vapor. In our experiment, an alphabet image was transferred from a coupling field to a probe field based on the CPT effect. Furthermore, we demonstrated that the cloned probe field carrying the image transmitted without usual diffraction. To our best knowledge, there is no any such an experimental report about images cloning beyond diffraction. We believe this mechanism based on CPT definitely has important applications in image metrology, image processing and biological imaging. © 2013 Optical Society of America.

Arbitrary images with finite size encoded in a light are subjected to the diffraction when it propagates in the free space or media. The reason is that the finite size of arbitrary images can be considered as a group of different plane-wave components and each component acquires different phase in propagation. Improving the resolution of an arbitrary image is a fundamental problem in imaging processing, and controlling optically an arbitrary image without diffraction is definitely very important for biology imaging [1], medical imaging [2] and so on. A way to optically realize it is applying a control field to induce a waveguide to affect the propagation of weak probe, which can result in the self-focusing (or self-defocusing) [3]. The physics mechanism is that the cross-phase modulation induces the different phase shift to the probe field in its propagation. Such cross-phase modulation techniques that induce inhomogeneous index refraction have been suggested in some works, including ones based on electromagnetically induced transparency [4-6], coherent population trapping (CPT) [7-10]. Although Ref. 11 reported that the structure profile of coupling field can be transferred to the probe field due to CPT, there is no any discussion about the propagation beyond diffraction of the probe in the free space or other media. Refs. 12 and 13 studied the propagation properties of the imprinted probe field and predicted that the probe field could propagate beyond diffraction limits in an induced media. Ref. [13] demonstrated that an arbitrary image could propagate beyond diffraction limits due to the compensated different phases contributed by a controlling field. Very recently, Ref. 14 predicted that an arbitrary image of control field could be cloned into the probe beam beyond diffraction. In that work, the spatial-independent phase shift of each plane-wave component of probe field can be controlled by the control field, by which the different phase shift of probe in propagation can be compensated. This is significantly different from the underlying physics of Ref. [15, 16]: where the phase difference acquired during propagation for each plane-wave component of the image is exactly compensated by an additional phase shift induced by the atomic motion, thus leading to the elimination of diffraction.

However, controlling an arbitrary image beyond usual diffraction by using atomic gas is a challenging work. In this paper, we reported on an experimental realization of a special image cloning beyond diffraction through CPT effect in a hot rubidium vapor. In our experiment, an image is imprinted on the coupling field, and then the image is cloned to the probe when both fields propagate through the hot atomic vapor. The cloned probe field carrying the image transmits without the usual diffraction, in which the image becomes much clear with respect to the diffraction.

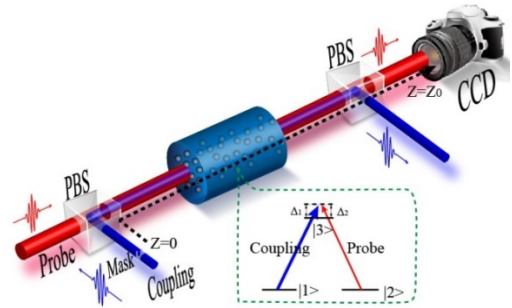

Fig. 1. (Color online) Simplified experimental diagram. The red line was the probe field and the blue line represented the coupling field. The mask was inserted into the optical route of coupling beam. The probe and coupling fields were monitored by a common camera. The experimental energy level configuration was shown in the dashed box.

In our experiments, we used a Λ-type CPT configuration (see Fig. 1), consisting of two ground states |1> and |2>, and one excited state |3>. The ground states were given by the Zeeman-degenerate levels of $^{85}$Rb atom ($5S_{1/2}$, F=3), the excited state corresponded to the level of $5P_{1/2}$, F=3. The coupling and probe fields had the same wavelength of 795 nm, the probe coupled the transition from state |2> to state |3> with a blue-detuned of $\Delta_2$, and the coupling field coupled the atomic transition of $5S_{1/2}$(F=3) -> $5P_{1/2}$(F'=3) with a blue-detuned of $\Delta_1$. The coupling was imprinted a real image by a mask of a standard resolution chart (USAF target). The polarizations of coupling and probe were orthogonal, and were combined into a beam through a 5-cm long vapor cell containing $^{85}$Rb atoms by using a polarization beam splitter. The output of coupling and probe fields were separated by using another polarization beam splitter. The structure of coupling and probe fields were monitored by a common camera. The specific parameters of our experiment were recorded as following: The powers of the probe and coupling laser beam were 1.4 mW and 1.5 mW, corresponding to the Rabi frequencies of $g=8.4\gamma$ and $G=29\gamma$, where $\gamma$ is the decay rate of the upper level |3>. The detuning of probe and coupling laser beam was $\Delta_1$=361 MHz and $\Delta_2$=375 MHz respectively. The diameter of the probe beam was 5 mm, and the coupling's was 1.5 mm, so the probe field completely covered the coupling field. The temperature of vapor cell was heated to be 76°, the atomic density of cell is about $2.5\times10^{12}$ cm$^{-3}$.

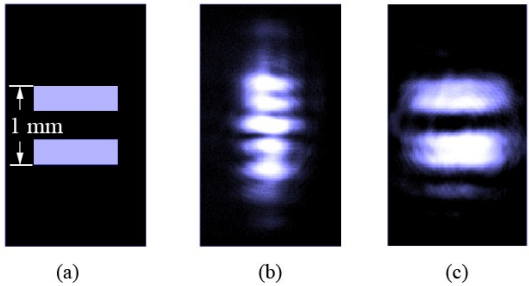

Fig. 2. (Color online) (a) the two-slit structure imprinted on the coupling laser beam. (b) the diffracted image obtained at z= $z_0$. (c) the cloned probe beam monitored at z= $z_0$. $z_0$=300mm.

In our experiment, we made the coupling beam carry an image of a two-slit structure, and then monitored its spatial information at z=$z_0$=300 mm by using a camera, where z was defined as the distance from the mask to CCD camera. There were clear interference fringes due to its diffraction when the coupling beam propagated in the free space (Here we made the frequency of the coupling field far-detuned with the atomic transitions, so the interaction between the laser and the atoms can be ignored).

Then we input the probe field into the vapor cell along the coupling beam. The obtained spatial structure of probe field at z=$z_0$=300 mm was shown in Fig. 2 (c). The structure of probe at z=$z_0$ was the same as the coupling field at z=0 with small blurry. Obviously, the probes kept the main characters of the input images and showed the good similarities to the input coupling field. This image was beyond the usual diffraction (Fig. 2(c)) compared with the coupling field z=$z_0$ where the strong distortions of spatial information appeared due to the diffraction (Fig. 2(b)). The small blurry appeared in Fig. 2(c) was still mainly due to the residual diffraction: there existed free space between the front surface of vapor cell and the mask (~45 mm), the diffraction during this distance cannot be controlled via the atomic vapor in our experiment. Another reason was from the unbalanced heating to cell in our experiment, which causes the small modulation on the index refraction, and makes the cloned image somehow blurry. Therefore if could improve the heater system, the quality of cloned image could be improved further. As it is claimed in theoretical paper Ref. 14, the sharpness of the cloned probe image can be increased by a factor 2 as compared to initial feature of the control image. Therefore, one may adjust the position of the CCD to achieve the sharper cloned image. In our system, we didn't use the 4-f image system to solve the diffraction between the vapor cell and the mask in order to directly illustrate the mechanism of cloning beyond the usual diffraction. In our experiment, the cloning effect beyond the usual diffraction was directly obtained without any 4-f image systems, which was straightforward and more convincing.

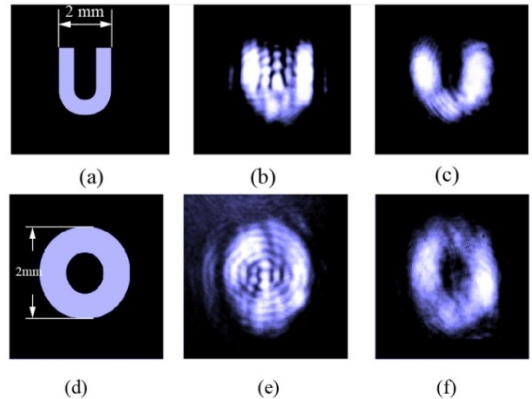

Fig. 3. (Color online) (a) The alphabet U imprinted on the coupling laser beam. (b)The diffracted image at z=$z_0$. (c) The cloned probe beam at z= $z_0$. (d-f) corresponded to the similar experimental results with the alphabet O. $z_0$=300mm.

Next, we repeated the experiments with two other images: alphabets U and O. We made the coupling beam carrying these two images respectively, and redid the experiments as before. Fig. 3 (a) and

Fig. 3 (d) were the imprinted images onto the coupling field. The propagated images of coupling field at z= $z_0$ were shown in Fig. 3 (b) and Fig. 3 (e). The cloned probe images at z=$z_0$ were recorded shown in Fig. 3 (c) and Fig. 3 (f). Obviously, the probes also kept the main characters of the input images and showed the good similarities to the input coupling field except some small blurry. In this process, the transmission intensity of cloned image was ~40 μW.

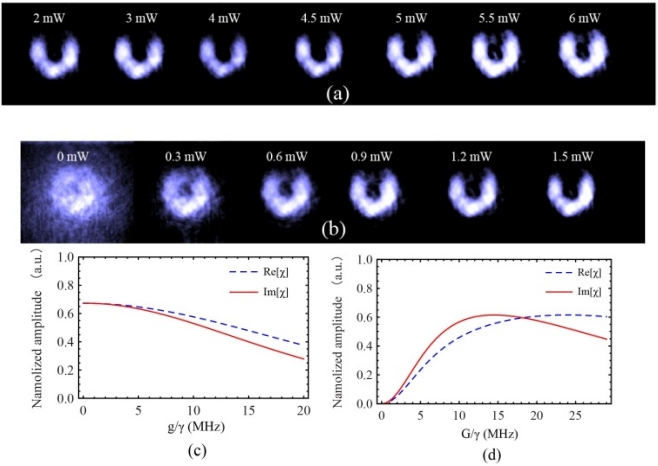

Fig. 4 (a) The closed image beyond the usual diffraction was shown against the power of probe field. The power of coupling field was about 1.5 mW. (b) The closed image beyond diffraction was shown against the power of coupling field. The power of probe field was about 4 mW. In these figures, the power of probe field 2~6 mW corresponds to the rabi frequency $10\gamma$~$17.4\gamma$; the power of coupling field 0~1.5 mW corresponds to the rabi frequency $0\gamma$~$29\gamma$. (c) and (d) were the calculated susceptibility as the function of rabi frequencies $g$ and $G$.

In the following, we checked the cloned image against some experimental parameters. Firstly, we set the power of coupling field to be 1.5 mW and monitored the cloned image against the power of probe field. The results were given by Fig. 4(a). And then, we modulated the power of coupling field to find the relation with the cloned image. The experimental results were shown by Fig. 4(b). With the increment of the coupling power, the effect of cloned images beyond diffraction became better. It seems that there was no strong relation between the power of probe field and the quality of the cloned image. This phenomenon was because the susceptibility became small, which results in the small induced phase and weak absorption. According to Ref. [14], we also derived the density-matrix equation and obtained the susceptibility of $\chi=\chi_{32}$ which was the function of detuning $\Delta_1$, $\Delta_2$; atomic density; rabi frequencies $G$, $g$ and decay rate $\gamma$. We characterized the real/imaginary part of the susceptibility $\chi$ against the rabi frequency $g$ shown in Fig. 4(c), where the rabi frequency of coupling field was set to be $G=29\gamma$ and the effective atomic density was $1.0\times10^{12}$ cm$^{-3}$. The curve in Fig. 4(c) illustrated the real/imaginary of susceptibility $\chi$ slowly varied with the different rabi frequency $g$. This point was consistent with our experimental results shown in Fig. 4(a). In the range of $G=10\gamma$~$0\gamma$, the susceptibility was directly attenuated from 0.5 to 0. From our experimental results in Fig. 4(b), the absorption of probe field 0.9 mW~ 0mW became small and the cloning effect became indistinct. This phenomenon was because the real/imaginary of the susceptibility $\chi$ became small, which results in the small induced phase and weak absorption.

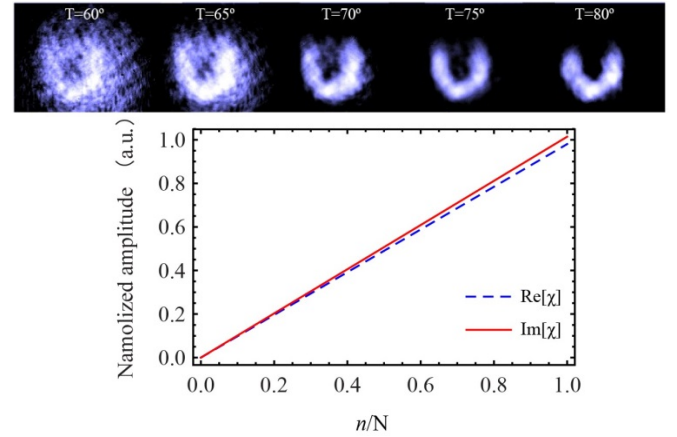

Fig. 5 The cloned image was against the temperature of the vapor cell (upper figure). The normalized susceptibility against the atomic density $n$. (down figure)

At last, we checked the cloning effect against the atomic density of the $^{85}$Rb in the cell. We used a heater to heat the vapor cell to change the atomic density. The results were shown by Fig. 5 (upper figure). It was shown that the cloned image became unclear with the decrement of the temperature of the vapor cell. This was because the cloning effect beyond diffraction needed more atoms, in such way the medium could be modulated with spatial index of refraction. The calculated curve could illustrate this reason: the real/part of susceptibility of $\chi_{32}$ linearly decreased with the decrement of atomic density which was shown in Fig. 5 (down figure) below where N=$1.0\times10^{12}$ cm$^{-3}$.

In conclusion, we reported on an experiment about cloning an image through CPT effect and its diffraction effects. With spatially dependent control filed, the medium has spatial dependent index of refraction and can effectively transfer the image of the control to the probe beam. The spatial independent phase shift of each plane-wave component of probe field can be controlled by the control field, by which the different phase shift of probe in propagation can be compensated, thus the probe beam transmits without the usual diffraction.

We also considered these effects in the different parameters such as: the power of coupling and probe fields and the temperature of vapor cell. Such experimental results clearly showed some interesting properties of CPT on image transfers, and we believe this effect definitely has important applications in image metrology, image processing and biological imaging etc.

**Acknowledgments**

We thank Dr. Tarak N. Dey and Dr. Chang-ling Zou for helpful discussions. This work was supported by the National Natural Science Foundation of China (Grant Nos. 11174271, 61275115, 10874171), the National Fundamental Research Program of China (Grant No. 2011CB00200), the Youth Innovation Fund from USTC (Grant No. ZC 9850320804), and the Innovation Fund from CAS, Program for NCET.